\documentclass[journal=jpcbfk,manuscript=letter,layout=twocolumn]{achemso}
\usepackage[utf8]{inputenc}
\usepackage{siunitx}
\usepackage{amsmath}
\usepackage{graphicx}
\usepackage{amssymb}
\usepackage[superscript,biblabel]{cite}
\usepackage{rotating}
\usepackage{xcolor}

\title{Sum-Frequency Signals in 2D-Terahertz-Terahertz-Raman Spectroscopy}
\author{Griffin Mead, Haw-Wei Lin, Ioan-Bogdan Magdău, Thomas F. Miller III \& Geoffrey A. Blake}
\date{\today}
\email{gab@gps.caltech.edu}
\begin{document}

\maketitle

\begin{figure}[t]
    \centering
    \includegraphics[width=0.5\textwidth]{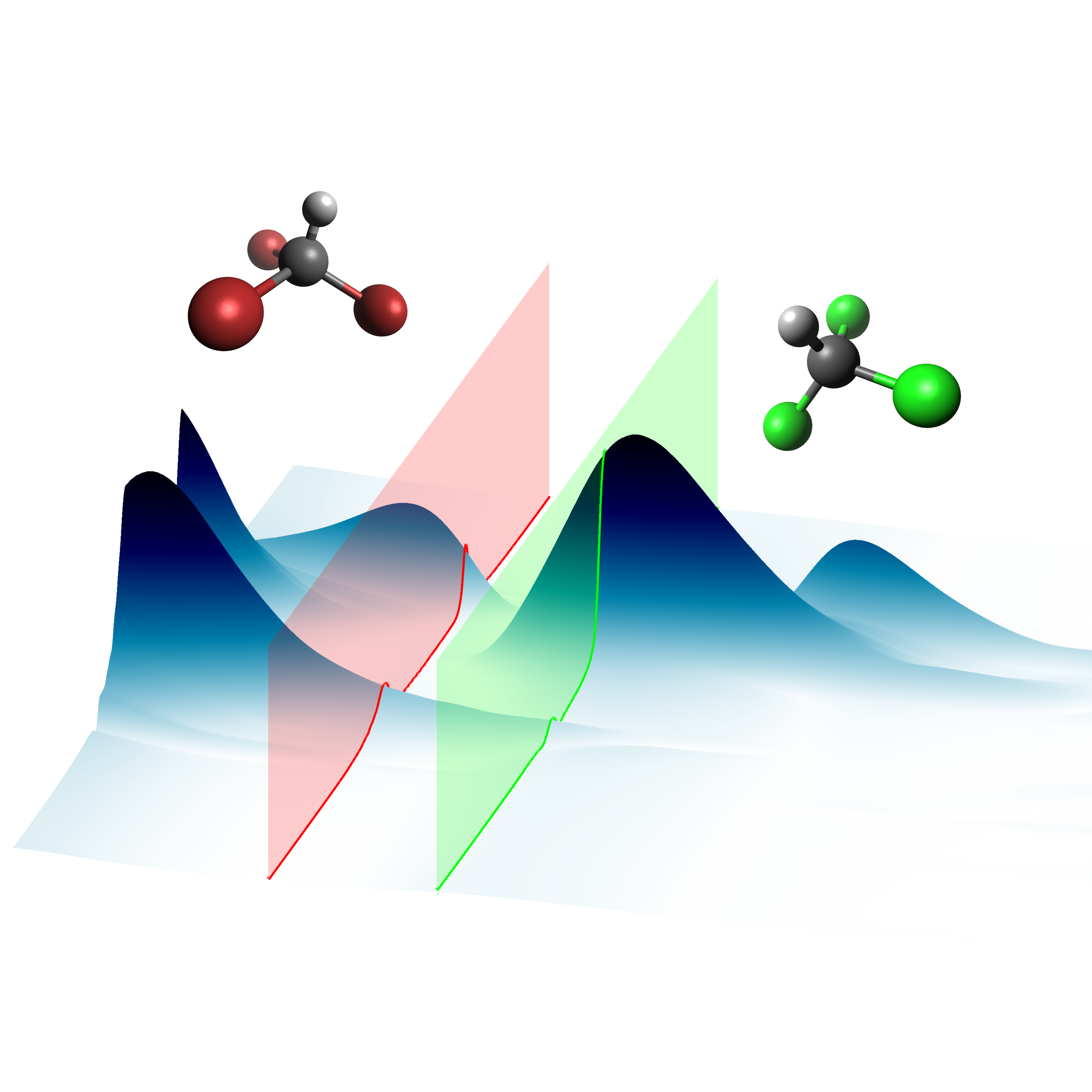}
    \caption{Table of contents figure}
    \label{fig:fig0}
\end{figure}

\section{Abstract}
We demonstrate that halogenated methane 2D-Terahertz Terahertz Raman (2D-TTR) spectra are determined by the complicated structure of the instrument response function (IRF) along $f_1$ and by the molecular coherences along $f_2$. Experimental improvements have helped increase the resolution and dynamic range of the measurements, including accurate THz pulse shape characterization. Sum-frequency excitations convolved with the IRF are found to quantitatively reproduce the 2D-TTR signal. A new Reduced Density Matrix model which incorporates sum-frequency pathways, with linear and harmonic operators fully supports this (re)interpretation of the 2D-TTR spectra.

\section{Main text}
Observing interactions within the low-frequency, thermally populated continuum of bath states is critical to developing a molecular understanding of liquid dynamics at room temperature. This energy regime is predominantly characterized by broad inter-molecular modes with short coherence times ($\sim$100s fs) which complicate the measurement and interpretation of potential energy, dipole and polarizability surfaces. One exception to this general observation are the intra-molecular vibrational modes of the halogenated methane (HM) family of liquids, whose well-defined coherent vibrational signals have long been observed in optical Kerr effect (OKE) experiments.\citep{McMorrow1988, Cho1993} 

Multidimensional time-resolved spectroscopy methods seek to disentangle these ambiguous spectra by introducing an additional time delay which separates dynamics along a second axis. The 5th order Raman technique\citep{tokmakoff1997} extends OKE to two dimensions and provides information on electrical and mechanical anharmonicities of the liquid, but practical implementation of this method is quite challenging.\citep{Blank1999, kubarych2002diffractive} A trio of 3$^{rd}$ order terahertz-Raman hybrid spectroscopies have been proposed as alternatives to 5$^{th}$ order Raman spectroscopy that avoid some technical challenges inherent to 5$^{th}$ order spectroscopy.\citep{Hattori2010,Ikeda2015,savolainen2013two-dimensional,shalit2017terahertz,hamm2017perspective} However, new challenges emerge in the hybrid techniques, especially compared to the more common 2D-infrared (2D-IR) spectroscopy. First, there are no commercially available dispersive THz spectrometers with adequate sensitivity to directly detect the emitted THz signal in THz-Raman-THz (2D-TRT) and Raman-THz-THz (2D-RTT) measurements. Instead, the 2D-TRT/RTT techniques have used time-domain electro-optic sampling to capture the faint THz emission.\citep{Ciardi2019} 2D-TTR avoids this step by using a Raman probe pulse which generates an easily detected near-IR signal photon. In all the cases of hybrid THz-Raman spectroscopies, the poorly defined THz wave vector precludes a phase-matching box-CARS style geometry that could be used to discriminate between signals originating from different quantum mechanical coherence pathways. 

With 2D-TTR spectroscopy, complex spectra have been observed in several halogenated methane (HM) liquids, and were interpreted as signatures of coherent energy transfer pathways between intra-molecular vibrational modes.\citep{Allodi2015,Finneran2016c, finneran20172d, Magdau2019}. A thorough re-investigation of two HM liquids -- bromoform (CHBr$_{3}$) and chloroform (CHCl$_{3}$) -- casts doubt on this original interpretation. Our new investigation is enabled by the development of a single-shot 2D-TTR spectrometer\citep{Mead2019} which records tens of picoseconds of molecular dynamics in a single acquisition. From the order of magnitude speed-up, the new technique provides substantially higher signal-to-noise data which has allowed a much larger region of the molecular response to be measured, and at finer resolution. 

We demonstrate through experiment, models, and theoretical simulations that the features observed in the HM 2D-TTR spectra arise from convolutions between the instrument response function (IRF) and linear interactions with the molecular polarizability operator $\mathit{\Pi}$. This interaction requires a scattering with two instantaneous THz photons, and is therefore referred to throughout the text as a sum-frequency (SF) excitation process. (Very recent experimental and theoretical works have also observed efficient phonon excitation through the same linear-$\mathit{\Pi}$ interaction with two THz field interactions.\citep{Maehrlein2017, Juraschek2018, Shishkov2019}) Resonant nonlinear interactions with the transition dipole operator $\mathit{M}$, while also in principal weakly allowed, are not detected.

\begin{figure}[t]
    \centering
    \includegraphics[width=0.5\textwidth]{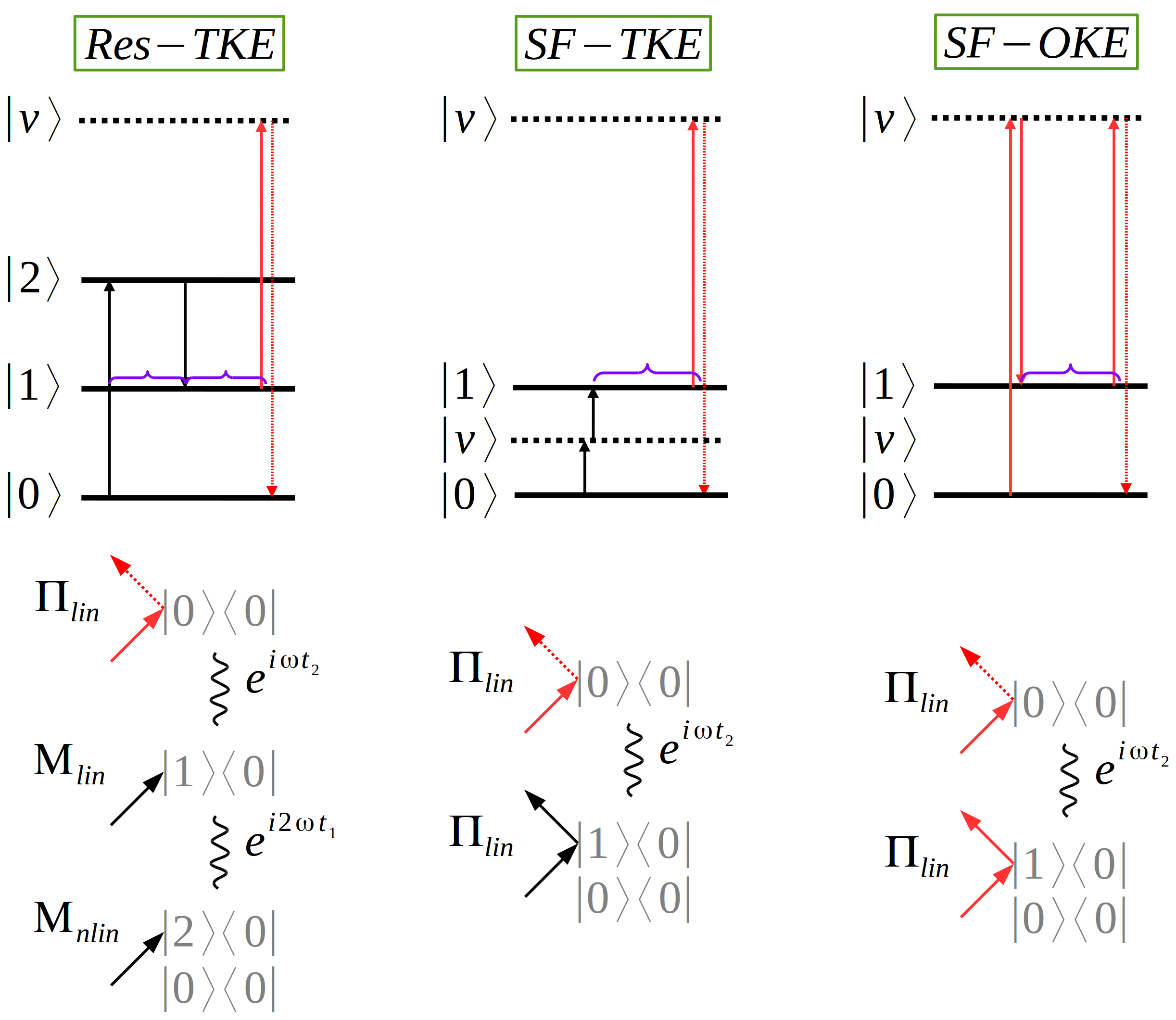}
    \caption{A resonant TTR signal requires dipole nonlinearities to excite a vibrational coherence - a representative process is illustrated in the Res-TKE (resonant terahertz Kerr effect) ladder and Feynman diagrams. In contrast, a sum-frequency excited molecular coherence is produced through interactions linear in the polarizability operator. The virtual state in SF-TKE is short-lived, and therefore the signal is highly dependent upon overlap between the two pump pulses. The familiar optical Kerr effect (SF-OKE) illustrates fundamental similarities with SF-TKE.}
    \label{fig:fig1}
\end{figure}

We begin by re-examining the relative importance of the $\mathit{M}$ (resonant) and $\mathit{\Pi}$ (sum-frequency) excitation pathways in HM vibrational modes. Ladder diagrams in Fig. \ref{fig:fig1} illustrate the two competing pathways as well as the OKE process, which is analogous to SF-TKE. In order to observe the desired nonlinear THz signal, the resonant pathway must have a larger or (at least) comparable magnitude with the sum-frequency pathway. This is a difficult condition to satisfy in HMs since the resonant process is nonlinear with respect to $\mathit{M}$ while the sum-frequency pathway is linear in $\mathit{\Pi}$.

Sum-frequency and resonant excitation pathways have distinct $t_1$ responses. A clear sign of resonant $M$ interactions is a prolonged vibrational response along $t_1$ which arises from the generation of a vibrational coherence during the first THz field interaction. From 2D-TTR measurements, molecular coherences extending in excess of 5 ps along $t_{2}$ have been observed in HMs, suggesting a resonant signal should have a commensurate lifetime along $t_{1}$. In contrast, sum-frequency excitation cannot directly generate coherent states through a single field interaction, but instead require two instantaneous interactions. In this case, the extent of a molecular response along $t_1$ will be determined by the duration of temporal overlap of the two THz electric field waveforms.

\begin{figure*}[t]
    \centering
    \includegraphics[width=1.0\textwidth]{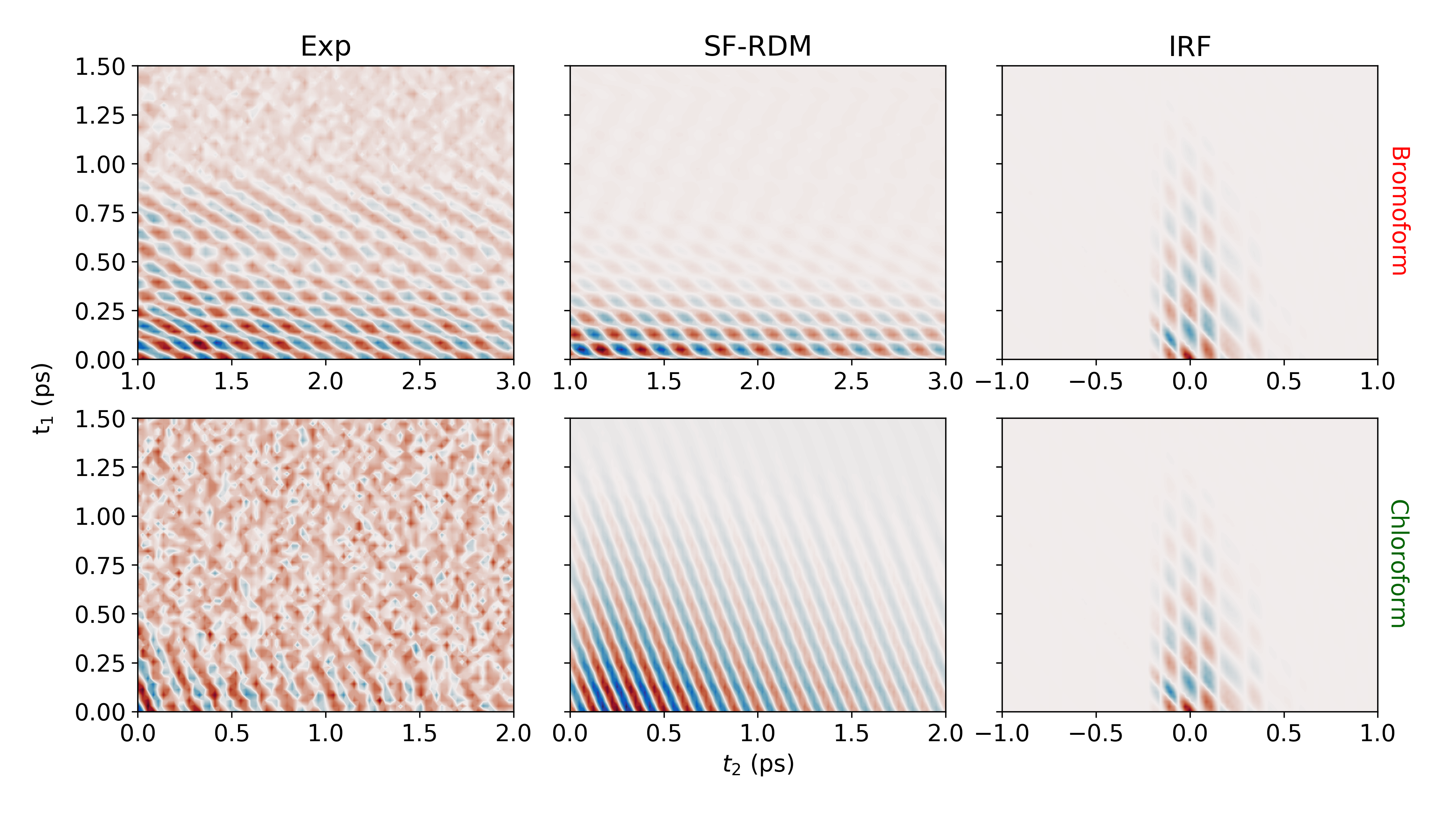}
    \caption{Top and bottom rows compare the experimental (Exp) bromoform and chloroform time-domain data to the SF-RDM models. The calculated IRF, whose THz electric fields are used as inputs to the SF-RDM model, is shown for reference (right column). Identical pulse shapes and IRFs are used for both SF-RDM models of bromoform and chloroform.}
    \label{fig:fig2}
\end{figure*}

In Fig. \ref{fig:fig2} time-domain bromoform and chloroform measurements recorded under identical experimental conditions are shown. The initial key observation is that while the $t_2$ response is long-lived, the response along $t_1$ never extends past the region of THz field overlap. Both bromoform's and chloroform's vibrational coherent responses are therefore far more consistent with a SF excitation mechanism than a resonant process.

The different bandwidth requirements of the two processes provides a second argument supporting SF excitation as the dominant pathway. In both SF and resonant 2D-TTR pathways, a vibrational mode must begin and end the measurement in a population state. In addition, the Raman probe interaction only changes the vibrational quanta by $\pm 1$. If a $\mathit{M}$ non-linearity is present, one of the THz field interactions must produce either a zero-quanta or two-quanta excitation, which in the latter case necessitates a bandwidth spanning $\geq 2 \omega$.\citep{Sidler2019} (No evidence for zero-quanta transitions in HM vibrational modes have been observed.) Again, this is quite different from 2D-IR spectroscopies. SF excitation, on the other hand, progresses with a bandwidth $\approx \omega/2$. The experimental THz field bandwidth spans 1-5 THz, and therefore the 2D-TTR experiment lacks the necessary frequency content to produce vibrational coherences arising from $\mathit{M}$ non-linearities in the molecular Hamiltonian.

Given that sum-frequency excitation is the predominant source of the signals observed in 2D-TTR spectroscopy of HMs, a key mystery becomes how this mechanism, whose instantaneous nature precludes separating the two THz field interactions in time, can nonetheless produce a signal which varies along $t_1$? We interrogate the origins of this complex $t_1$ response by considering how the observed signal $S(t_1, t_2)$ depends upon the IRF $I(t_1, t_2)$ (Eq. \ref{eq:eq1}). In a 2D-TTR experiment, two orthogonally polarized THz fields ($\vec{x}, \vec{y}$ in the laboratory frame) create a birefringent response within the room-temperature HM liquid sample. A $\vec{x}$ polarized Raman probe  scatters off this birefringence, producing a $\vec{y}$ polarized signal field that is selectively isolated through an analyzing polarizer and differential chopping. The 2D-TTR signal is proportional to the anisotropic third-order molecular response function $R_{xyxy}^{(3)}(t_1, t_2)$, and contains information on the molecular orientational and vibrational correlation functions of the system. During the measurement process, this response is inevitably convolved with the experimental IRF, which in 2D-TTR is determined by the product of the two THz electric fields.

\begin{equation}
    S(t_{1}, t_{2}) = I(t_{1}, t_{2}) \circledast R^{(3)}_{xyxy}(t_{1}, t_{2})
    \label{eq:eq1}
\end{equation}

Through the convolution theorem, the time-domain convolution becomes a multiplication between the IRF spectral power and the HM molecular response function upon transformation to the frequency domain. 

\begin{equation}
    \tilde{S}(f_{1}, f_{2}) = \tilde{I}(f_{1}, f_{2}) \cdot \tilde{R}^{(3)}_{xyxy}(f_{1}, f_{2})
    \label{eq:eq1}
\end{equation}

We study the impact of IRF convolution in two ways. First, we generate a time-domain model IRF (Fig. \ref{fig:fig2}) using model THz field profiles (Fig. \ref{fig:fig5}) that closely resemble experimental pulse shapes (see SI for details). An instantaneous SF process gives a molecular response which is a delta function in the time-domain ($t_1$) and a flat response in the corresponding frequency domain ($f_1$). In 2D-TTR, this amounts to a flat response along the $f_1$ axis, and a delta functions along the $f_2$ axis centered at the eigenmode frequencies of the molecular sample. Multiplication of this molecular response with the IRF yields the final measured signal. This results in simply selecting a slice of the IRF along $f_1$ at the eigenmode frequency. Using this simple model we find excellent agreement with the experimental spectra. 

\begin{figure}[t]
    \centering
    \includegraphics[width=0.45\textwidth]{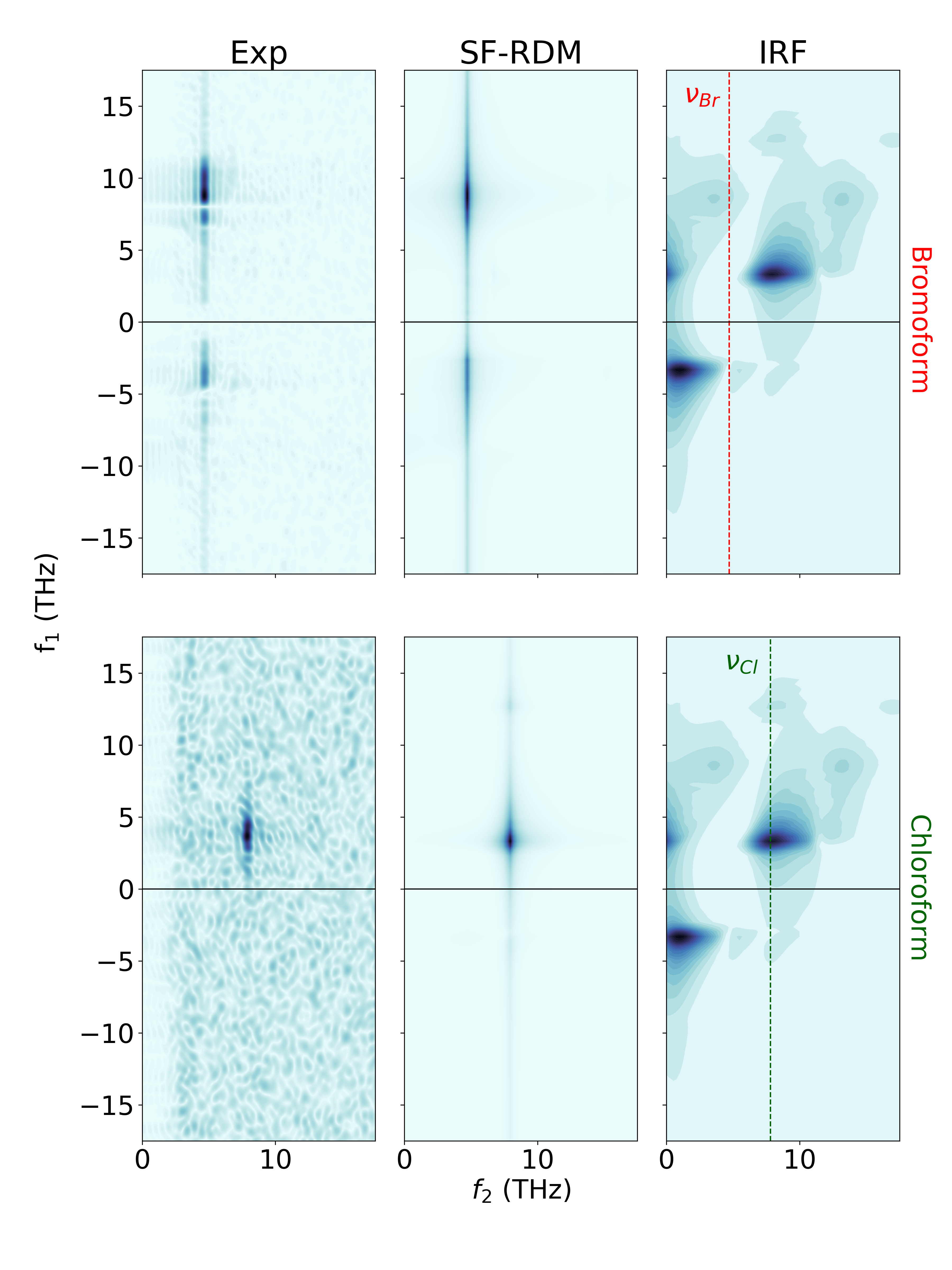}
    \caption{Top and bottom rows compare the experimental (Exp) bromoform and chloroform frequency-domain data to the SF-RDM model. The calculated IRF is shown for reference with vertical lines indicating where each HMs intramolecular vibrational mode samples the IRF. Note that the experimental and SF-RDM spectra are well matched, and arise from sampling the same IRF at different $f_{2}$ frequencies.}
    \label{fig:fig3}
\end{figure}

Second, we use the same THz field profiles that produce the IRF model as inputs to RDM simulations that consider sum-frequency excitation processes (SF-RDM). Again, we find near-quantitative agreement between the experimental data and theoretical simulations. Critically, no electrical or mechanical non-linearities are required to reach excellent agreement between the data and the IRF model/RDM simulations. The SF-RDM results precisely reproduce the experimental time-domain (Fig. \ref{fig:fig2}) and frequency-domain (Fig. \ref{fig:fig3}) responses, substantiating the claim that SF processes dominate the 2D-TTR response of HMs.

\begin{figure}[t]
    \centering
    \includegraphics[width=0.5\textwidth]{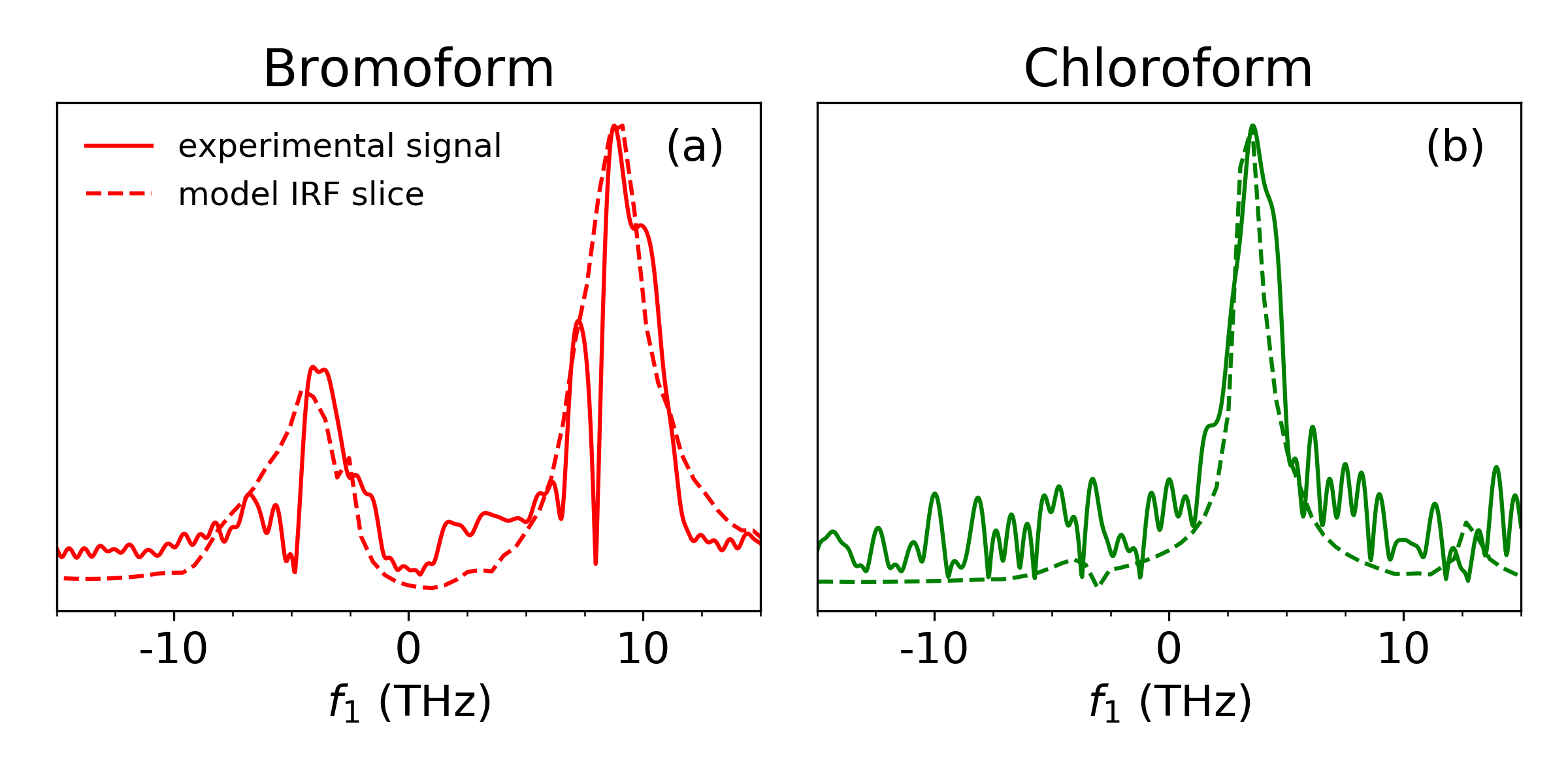}
    \caption{(a) Slices along $f_1$ at $f_2$=eigenmode of the IRF/RDM model and experimental response demonstrate the quality of fit for the bromoform data. Chloroform (b) is reproduced by slicing along the same IRF/SF-RDM model as shown in (a) at $f_2$=7.8 THz, instead of at bromoform's $f_2$=4.7 THz.}
    \label{fig:fig4}
\end{figure}

\begin{figure}
    \centering
    \includegraphics[width=0.5\textwidth]{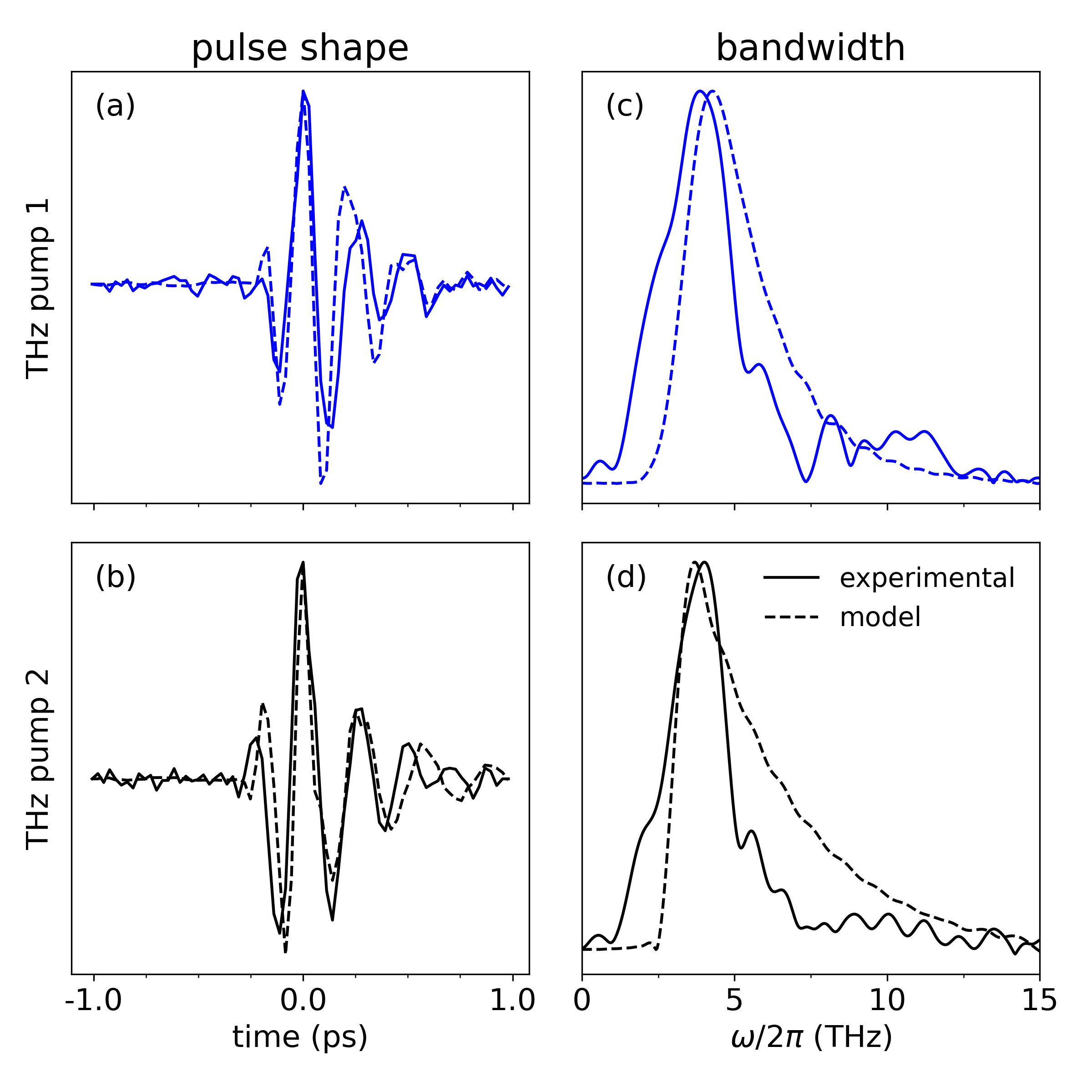}
    \caption{Comparison of the experimental and model THz pulse shapes (a,b) and corresponding bandwidths (c,d). See the SI for more detail on the optimization process used to obtain the model pulse shapes.}
    \label{fig:fig5}
\end{figure}

The agreement between experiment, SF-RDM, and model IRF are shown in Fig. \ref{fig:fig4}. While bromoform and chloroform have different intra-molecular vibrational energies, we reproduce both spectra by slicing the same model IRF at their respective eigenmode frequencies along $f_2$. Crucially, this model mimics the non-specific sum-frequency excitation of vibrational coherences by the THz electric field. Unlike previous interpretations, here we do not invoke Feynman diagram pathways involving multi-quanta transitions between several vibrational modes; instead, the experimental IRF filtered through a single SF excitation pathway explains our observations.

That a two-quanta sum-frequency excitation pathway is predominantly responsible for vibrational coherences observed in 2D-TTR is also supported by estimations of chloroform and bromoform's transition dipole moments. From THz-TDS measurements and FT-IR literature, the bromoform and chloroform $E$ mode's molar extinction coefficients were $\epsilon \approx1 M^{-1}cm^{-1}$, suggestive of a vanishingly small transition dipole moment. Ab initio calculations similarly arrive at transition dipole elements in the few milliDebye range. As a result, it would be very difficult to observe resonant excitation of these HM modes, even in the absence of interfering SF pathways. For comparison, 2D-IR spectroscopy on proteins is often performed by resonantly pumping the amide I stretch at $\sim50$ THz, which have $\epsilon\sim$200-400 M$^{-1}$cm$^{-1}$. Not only are these oscillators intrinsically orders of magnitude stronger than HM vibrational modes, but the IR excitation field's $\delta\omega/\omega$ is also substantially narrower, which helps to selectively and resonantly generate the desired coherences while suppressing any SF contribution. Common sources of high intensity, sub-ps THz pulses (organic emitters, LiNbO3, etc) all have $\delta\omega \approx \omega$ and thus both Res and SF pathways must be considered when analyzing responses in the overlapping pump field region.

Finally, we would like to note that our conclusions regarding the excitation mechanism of intra-molecular vibrational modes of HMs in 2D-TTR spectroscopy likely do not alter analyses performed on similar systems in the complementary 2D-TRT and 2D-RTT experiments.\citep{Ciardi2019} Those measurements attributed spectral features that remained post-deconvolution to couplings between a Raman-excited vibrational coherence and a resonant one-quanta interaction with bath modes of the liquid. Our conclusions are consistent with their observation that the IRF strongly determines the observed multi-dimensional experimental response.

In this work we provide extensive new experimental and theoretical data that leads to a simple reinterpretation of previous 2D-TTR measurements of HMs. With this new analysis, we explain the entire spectrum of both bromoform and chloroform through a convolution of the experiment's THz fields with the molecules' intra-molecular vibrational modes. No coherence pathways outside of the SF-TKE process in Fig. \ref{fig:fig1} are required. The new analysis is also fully consistent with the observed magnitudes of transition dipole moments and molecular polarizabilities. 

Moving forward, there are two keys lessons. First, large transition dipoles are crucial for performing truly resonant 2D-TTR experiments. Halogenated methanes unfortunately do not satisfy this requirement, and the bright, complex signals observed can easily be mis-attributed to resonant processes. Second, nearly transform-limited half-cycle THz fields would maximize the field strengths achievable, greatly simplify the experimental IRF structure, and reduce ambiguities in the analysis of dynamics in molecular systems. 

\bibliography{main.bib}
\end{document}


\maketitle

\maketitle
\section*{2D-TTR Single shot spectrometer}
Greater temporal resolution and sensitivity compared to previous research was achieved by developing a single-shot 2D-TTR spectrometer. This technique eliminates scanning of the probe delay line, producing substantial reductions in experimental acquisition time. With the imaging and echelon combination used for this work, each camera acquisition captures 30 ps of molecular response along $t_{2}$ at 28 fs resolution. Equivalent resolution of 28 fs steps are recorded along $t_{1}$, while the overall temporal extent of a typical data set spans 30$\times$[5-10] ps ($t_{2} \times t_{1}$). This represents $\geq10\times$ greater temporal extent compared to previous stage-scan 2D-TTR measurements\citep{finneran20172d}. Typical acquisition times ranged from a few hours to overnight (For comparison, previous stage-scan experiments required 48 hrs of averaging).  

\begin{figure}
    \centering
    \includegraphics[width=1.0\textwidth]{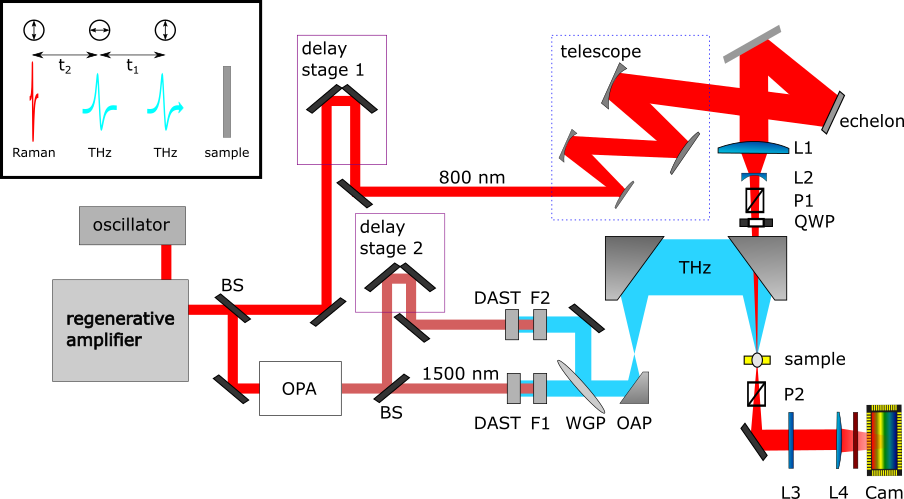}
    \caption{The optical path of the 2D-TTR single-shot spectrometer. Seen in an inset are the polarizations of the two THz pump and 800 nm probe fields, and the experimental time definitions.}
    \label{fig:SI_INSTRUMENT}
\end{figure}

We briefly discuss the overall experimental setup (Fig. \ref{fig:SI_INSTRUMENT}), and then provide specific details on the single-shot spectrometer's construction. A 1 kHz Coherent Legend HE regenerative amplifier, seeded with an 80 MHz Coherent Micra oscillator, is used to pump a Light Conversion TOPAS-C optical parametric amplifier (input 3.2 mJ, $\sim$30\% conversion efficiency signal+idler). The 500 $\mu$J 1.4 $\mu$m signal output is divided with a 50:50 beam-splitter before illuminating two 6 mm clear aperture DAST THz emitters (4-N,N-dimethylamino-4$'$-N$'$-methyl-stilbazolium tosylate, Swiss Terahertz). One OPA pump line is passed off a mechanical stepping delay stage which controls the relative time delay between the two pump fields. Additionally, one of the 1.4 $\mu$m pump line's polarization is rotated 90 degrees so that the THz emission from the two sources can be recombined and collimated with a wire-grid polarizer (WGP). After recombination, the THz light is filtered with 4 QMC 18 THz low-pass filters (2 after each crystal) to remove any residual optical bleed-through. The THz light is collected, collimated, and focused with a series of off-axis parabolic (OAP) mirrors before being focused onto the sample. The sample is held in a 1 mm path length Suprasil cuvette with a 4.6 mm diameter hole drilled in the front face. Over this hole a 5$\times$5 mm clear aperture, 1 $\mu$m thick silicon nitride membrane window (Norcada, QX10500F) is epoxied. This thin membrane minimizes dispersion differences between the THz pump and optical probe fields and also reduces the window response (previous works used 1 mm thick quartz or 300 $\mu$m thick diamond windows). 

The probe line beam conditioning for single-shot acquisition is unchanged from a previous study.\citep{Mead2019} Signal photons are detected on an Andor Zyla 5.5 MP sCMOS camera using a pair of crossed 800 nm polarizers with a 10,000:1 extinction ratio (Thorlabs LPVIS050-MP2). Because 2D-TTR uses two orthogonally polarized THz pump fields ($\vec{x}$,$\vec{y}$ polarized in lab coordinates) and crossed probe, signal polarizations ($\vec{x}$,$\vec{y}$ in lab coordinates), the detected signal is sensitive to the anisotropic third order molecular response function $R^{(3)}_{xyxy}(t_{1}, t_{2})$. 

 All data and experimental parameters were controlled from a MATLAB GUI purpose-built for the experiments. Within the GUI, number of acquisitions per average and number of averages are specified by the user. To capture a data set at a given $t_{1}$ time, typical acquisition settings for bromoform are to collect 5 averages with 2,000 shots per average, for a total acquisition time of $\sim$12 seconds, which includes 2 seconds of camera and computational overhead and the 10 seconds of data acquisition. The data set for each average is first collected as a 2000 $\times$ 1280 array, with dimensions specified by the number of laser shots captured (e.g. 2000) and the 1280 horizontal binned output pixels from the camera that capture the time-domain molecular response along $t_{2}$ at the fixed $t_{1}$ time. 

Next, we extract the signal from this data array by performing a FFT on each of the 1280 rows of data. Each row corresponds to a single point in time $t_2$ sampled by the probe pulse. These FFTs must be performed carefully to phase relationship of the two THz fields that produce the signal. To do so, we collect several other pieces of data while each average is acquired. First, outputs of the two optical choppers that modulate the THz pump fields are recorded on a DAQ card. Second, the camera output signal which confirms the start of a series acquisition is digitized by the same DAQ. From these three trigger signals (chopper A phase, chopper B phase, camera acquisition start), we can measure the signal phase of each average and correct for the phase offset. Doing so ensures that every average at every $t_{1}$ time is collected with a single consistent phase, which preserves the overall phase of the 2D signal.

The FFTs are performed by first multiplying each of the 1280 rows of data (each with length 2000 in this example) with an equal length sinusoidal waveform with the measured phase offset and a frequency equal to the difference between the two chopper frequencies (i.e. differential chopping). Multiplication shifts the desired differential signal to the DC position in the frequency domain FFT output. The DC response from all 1280 channels are then saved as a 1$\times$1280 data array for each average, and are co-added until all averages are collected. In essence, this approach creates a software equivalent to the standard digital lock-in amplifier typically used for collection of single channel spectroscopic data.

\section*{2D-TTR Sensitivity to IRF}

\begin{figure}
    \centering
    \includegraphics[width=1.0\textwidth]{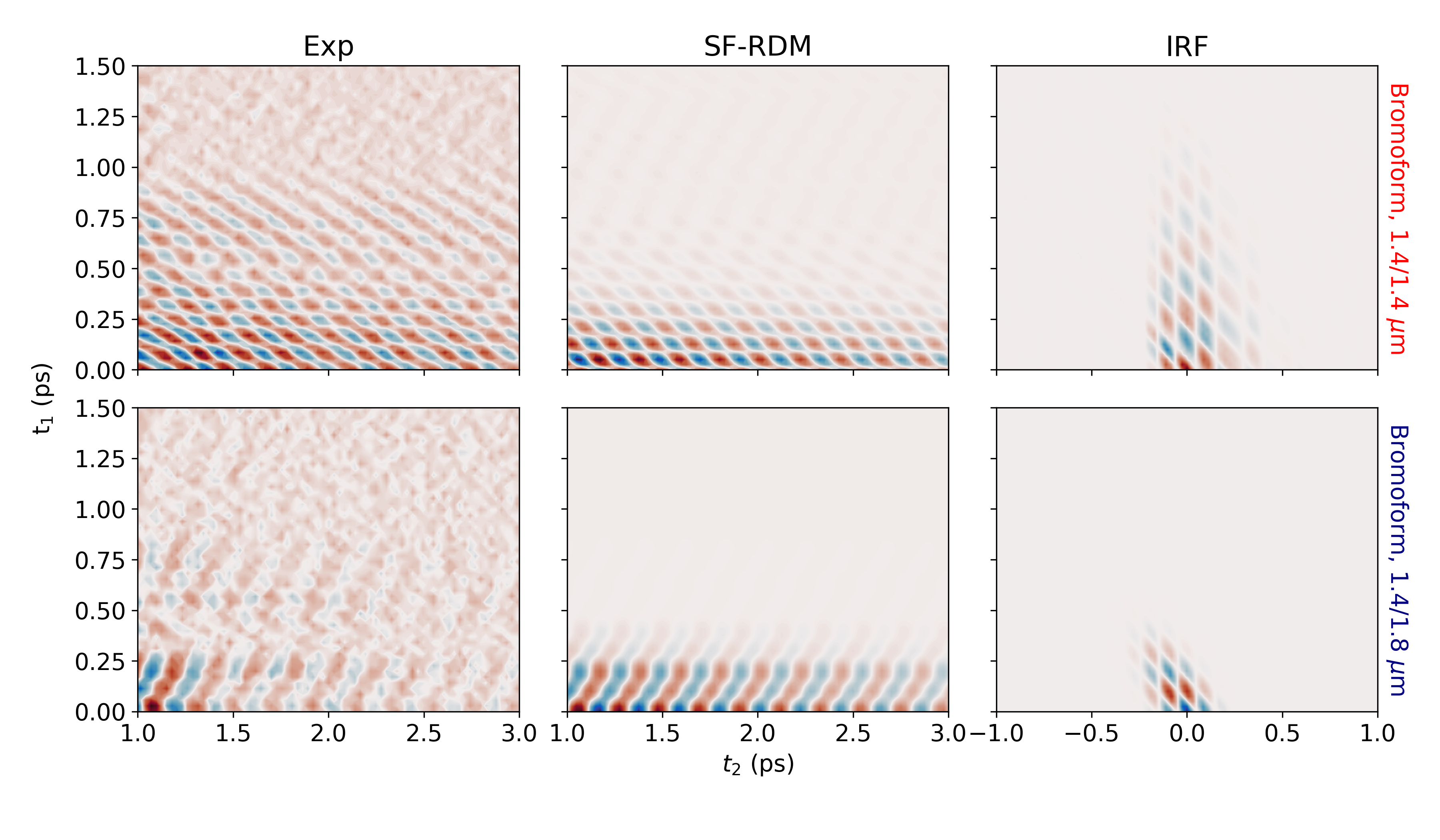}
    \caption{Top and bottom rows compare the experimental (Exp) bromoform time-domain data under two different THz pumping regimes. Slight changes to the emitted THz fields result in different time-domain IRFs. SF-RDM models using the THz electric fields that produce each IRF fully reproduce the experimental data.}
    \label{fig:BrBr_Time}
\end{figure}

\begin{figure}
    \centering
    \includegraphics[width=1.0\textwidth]{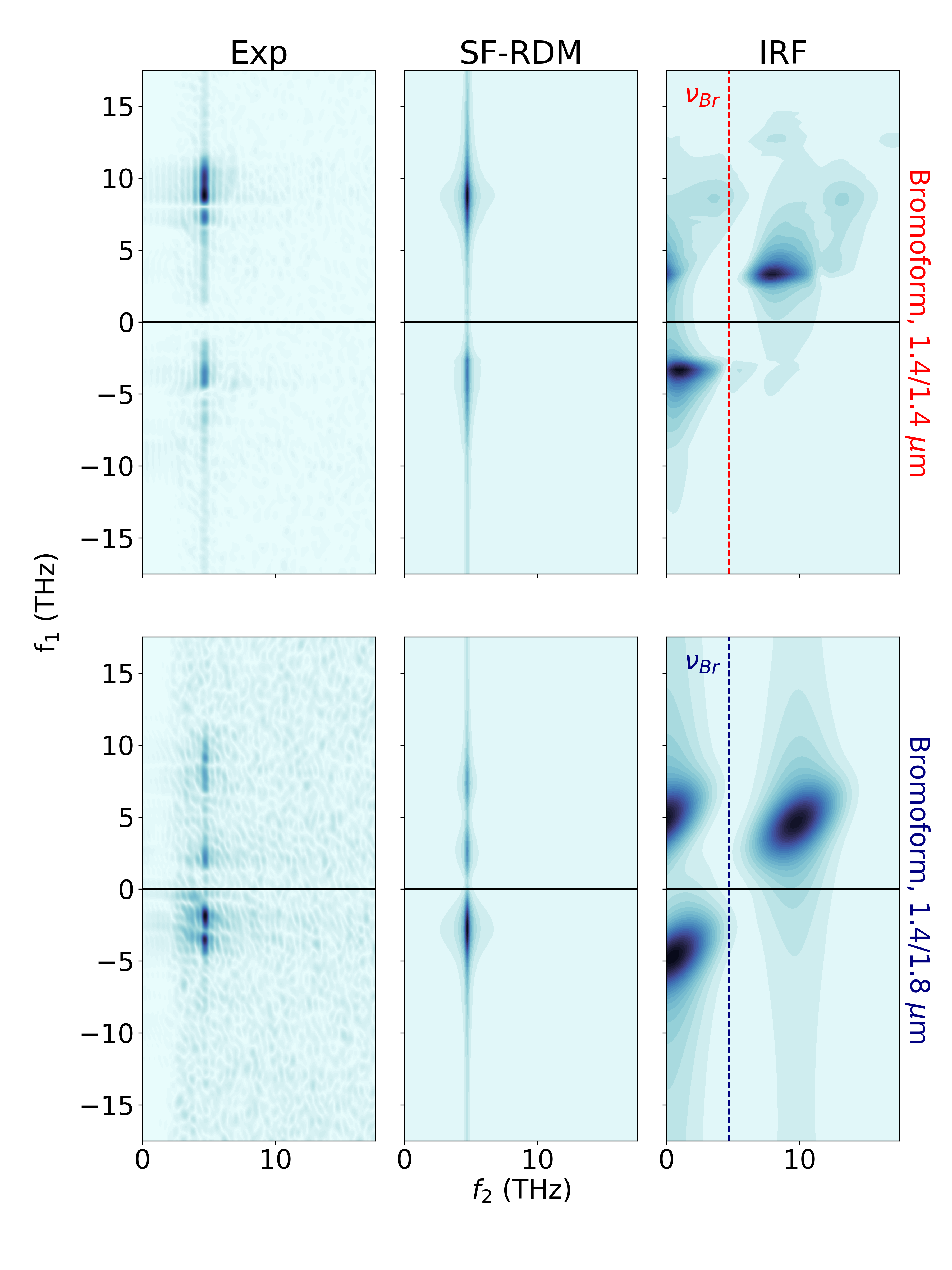}
    \caption{Top and bottom rows compare the experimental (Exp) bromoform frequency-domain data under two different THz pumping regimes. Slight changes to the emitted THz fields result in different frequency-domain IRFs. SF-RDM models with the THz electric fields that produce each IRF reproduce the experimental spectra.}
    \label{fig:BrBr_Freq}
\end{figure}

We provide an additional example of the sensitivity of bromoform's 2D-TTR spectrum to the instrument response function, which was first noticed as a result of changing the infrared wavelengths that pump the two THz emitters. Initial 2D-TTR experiments used the two orthogonally polarized signal (1.4 $\mu$m) and idler (1.8 $\mu$m) outputs of the TOPAS-C OPA were each used to pump one of the DAST crystals. However, we observed that the majority of the idler line's IR power was concentrated in a small $\sim$1 mm diameter point within the larger 8 mm diameter beam. As a result, this concentrated point of power was burning the DAST crystal face. To prevent damage to the emitter crystal, we split the signal beam into two halves, rotated one half's polarization to match the idler polarization, and recorded new 2D-TTR data of bromoform. Dramatic differences in bromoform's time (Fig. \ref{fig:BrBr_Time}) and frequency (Fig. \ref{fig:BrBr_Freq}) domain response were observed, drawing our attention to the importance of the IRF in 2D-TTR. SF-RDM analysis of the original 1.4/1.8 $\mu$m bromoform data yielded similarly excellent agreement between experiment and simulation.

\section*{Halogenated methane purification}
Bromoform was purchased from Sigma Aldrich. As received the liquid had a orange hue, indicating some degradation and contamination from water. A simple distillation under nitrogen at 150 degree Celsius was performed to purify the bromoform liquid. NMR analysis indicated complete removal of water. The purified sampled was stored in a round-bottom flask with dry sodium sulfate, under argon in a dark refrigerator. Chloroform was purchased from Sigma Aldrich and used as received.

\section*{TKE Response of HMs}
\begin{figure}[h]
    \centering
    \includegraphics[width=0.5\textwidth]{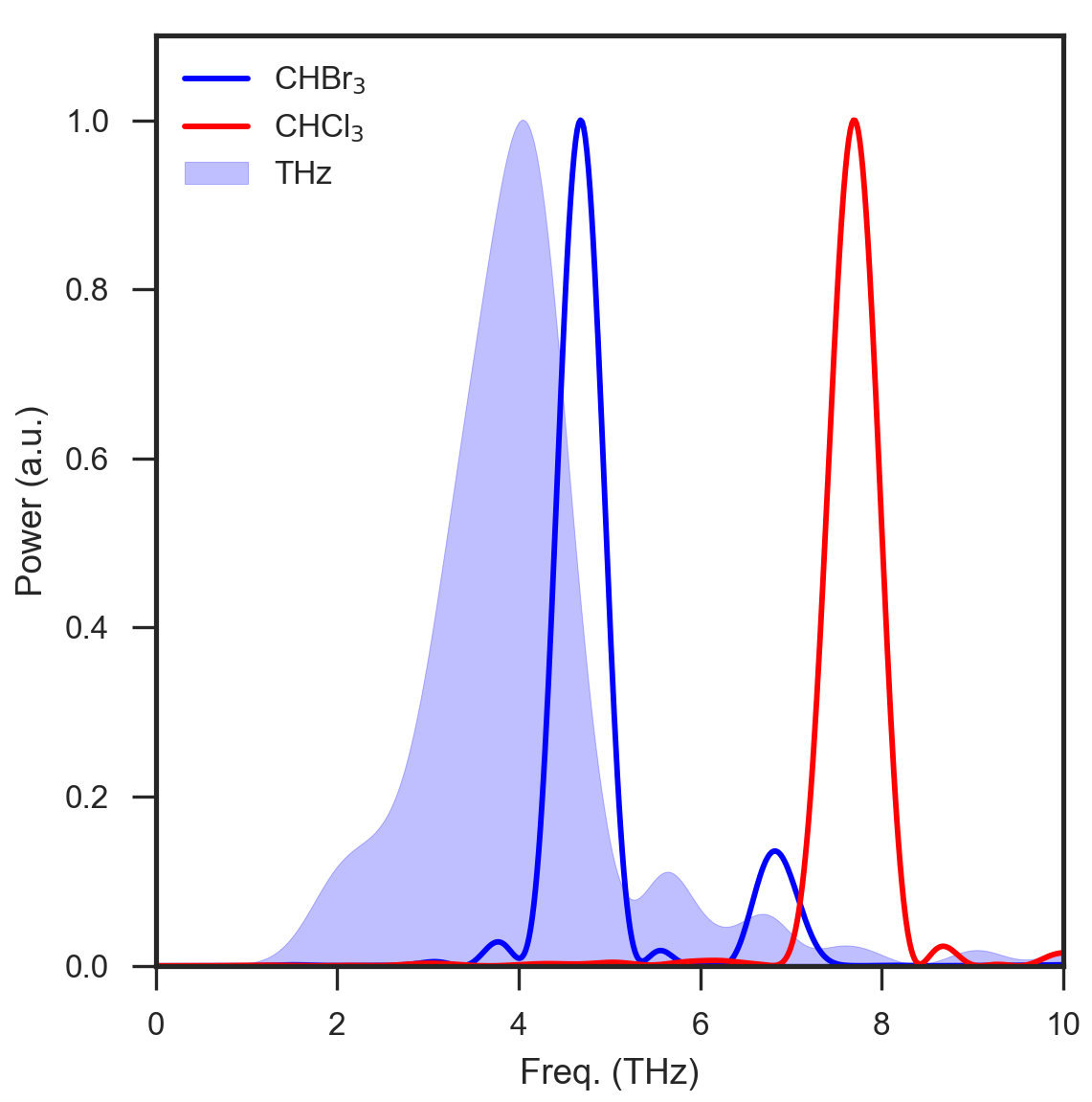}
    \caption{Bromoform and chloroform intramolecular vibrational responses are detected in 1D-TKE measurements. The THz bandwidth of the DAST emission as measured in diamond is shown in blue shading.}
    \label{fig:SI_BANDWIDTH}
\end{figure}

Terahertz Kerr effect measurements were taken of bromoform and chloroform, and the 2D-TTR response of diamond was measured to gauge the bandwidth of the THz pump fields.\citep{finneran20172d} As shown in Fig. \ref{fig:SI_BANDWIDTH}, the THz electric field bandwidth is centered at 4 THz. Bromoform's two Raman-active vibrational modes are observed at 4.7 and 6.6 THz, while the Raman-active vibrational mode in chloroform was observed at 7.8 THz. Note that the orientational (low-frequency) response of both liquids was removed prior to taking the FFT, and only the higher frequency spectral content arising from intra-molecular vibrations are seen in Fig. \ref{fig:SI_BANDWIDTH}. 

\subsection*{THz field modeling}
The frequency domain IRF slices at the eigenmode frequencies for halogenated methanes in this study are highly sensitive to small changes in the THz pump pulse shapes. This is especially true for bromoform, whose eigenmode frequency intersect a region of very low spectral power in the IRF. Near perfect fits to the experimental TTR spectra of HMs in both the time and frequency domains were obtained by using model THz pulse shapes, which closely resemble the experimental pulse shapes measured in diamond. We believe the use of these model THz pulse shapes is adequate here considering the chromatic nature of the THz focus and the S/N limitations of the diamond TTR spectra (due to the inherently small $\chi^{(3)}$ constant).

The model THz pulse shapes were obtained with the following process. First, the bandwidths of the two THz pulses were optimized to fit both experimental HM TTR slices at eigenmode frequencies of bromoform and chloroform. To avoid overfitting, a simple asymmetric Gaussian functional form was assumed for the bandwidth of the THz pulses to minimize the number of parameters. The time domain IRF is calculated as the product of the THz pulse shapes $$ I(t_1, t_2) = E_1(t_1+t_2)E_2(t_2) $$ and the Fourier transform of the IRF $\Tilde{I}(f_1, f_2)$ is $$ \Tilde{I}(f_1, f_2) = \Tilde{E_1}(f_1)\Tilde{E_2}(f_2-f_1)$$ where $\Tilde{E_1}$ and $\Tilde{E_2}$ are the bandwidths of the THz pulses. The frequency domain optimization yields only the amplitude spectra, and a phase spectra is still required to uniquely determine the time-domain representation. The hybrid input-output phase retrieval algorithm was used with the experimentally measured THz pulse shapes as targets. Convergence was generally obtained within 100 iterations. The resulting model THz pulses accurately reproduce the observed time and frequency domain results for both HMs, as shown in the main text. THz pulse shape measurements in general underestimate the available power at higher THz frequencies due to velocity mismatch between the probe (800 nm) and THz pulses, which further supports the increased power above 4 THz for the model pulse shapes.

\subsection*{SF-RDM model}

Sum-frequency pathways were not considered in our previous RDM model Hamiltonian \citep{Finneran2016c,finneran20172d,Magdau2019}: 
\begin{equation}
  H(t;t_1) = H_0 - M \cdot \left[ E_2(t-t_1) + E_1(t) \right]
\end{equation}
A new Hamiltonian was constructed here, which accounts for the SF process:
\begin{equation}
  H(t;t_1) = H_0 - \mathit{\Pi} \cdot \left[ E_2(t-t_1) + E_1(t) \right]^2
\end{equation}
Experimentally, differential chopping of the two THz fields automatically removes the single pulse contributions $\mathit{\Pi} \cdot E_2^2(t-t_1)$ and $\mathit{\Pi} \cdot E_1^2(t)$, so we are left with an effective Hamiltonian of the form:
\begin{equation}
  H(t;t_1) = H_0 - \mathit{\Pi} \cdot E_2(t-t_1)E_1(t)
\end{equation}
The time response is computed as described in our previous work\citep{Magdau2019}, but now it corresponds to an SF signal:
\begin{equation}
  S(t_2;t_1) = S_{SF}(t_2;t_1) = Tr(\mathit{\Pi}\cdot\rho(t_2;t_1))
\end{equation}
The complete signal $S$ is obtained when $E_1$ and $E_2$ are replaced with the fitted pulse shapes, while the molecular response $R$ when $E_1$ and $E_2$ are simple $\delta$-functions. All operators are kept linear and harmonic.

\bibliography{main.bib}